\documentclass[twocolumn]{autart}    
\makeatletter
\renewcommand\@biblabel[1]{}
\renewenvironment{thebibliography}[1]
     {\section*{\refname}%
      \@mkboth{\MakeUppercase\refname}{\MakeUppercase\refname}%
      \list{\@biblabel{\@arabic\c@enumiv}}%
           {\settowidth\labelwidth{\@biblabel{#1}}%
            \leftmargin\labelwidth
            \advance\leftmargin\labelsep
             \advance\leftmargin by 0.5cm%
            \itemindent -0.6cm
            \usecounter{enumiv}%
            \let\p@enumiv\@empty
            \renewcommand\theenumiv{\@arabic\c@enumiv}}%
      \sloppy
      \clubpenalty4000
      \@clubpenalty \clubpenalty
      \widowpenalty4000%
      \sfcode`\.\@m}
     {\def\@noitemerr
     {\@latex@warning{Empty `thebibliography' environment}}%
     \endlist}
\makeatother

\usepackage{graphicx}          
\usepackage{mathrsfs}
\usepackage{amsfonts,amssymb,color}

\usepackage{amsmath}

\begin{document}

\begin{frontmatter}

\title{On Potential Equations of Finite Games\thanksref{footnoteinfo}} 

\thanks[footnoteinfo]{This work is supported in part by National Natural Science
Foundation (NNSF) of China under Grant 11271194 and a project
funded by the Priority Academic Program Development of Jiangsu
Higher Education Institutions (PAPD). This paper was not presented
at any IFAC meeting. Corresponding author J.~Zhu. Tel.
+86-13851781823.}

\author{Xinyun Liu}\ead{liuxinyun1224@163.com},
\author{Jiandong Zhu}\ead{jiandongzhu@njnu.edu.cn}  

\address{School of Mathematical Sciences, Nanjing Normal
University, Nanjing, 210023, P.~R.~China}  


\begin{abstract}
In this paper, some new criteria for detecting whether a finite
game is potential are proposed by solving potential equations. The
verification equations with the minimal number for checking a
potential game are obtained for the first time. Some connections
between the potential equations and the existing characterizations
of potential games are established. It is revealed that a finite
game is potential if and only if its every bi-matrix sub-game is
potential.
\end{abstract}
\end{frontmatter}

\section{Introduction}
Game theory, the science of strategic decision making pioneered by
John von Neumann (see von Neumann, J. \& Morgenstern, O., 1953),
has wide real-world applications in many fields, including
economics, biology, computer science and engineering. The Nash
equilibrium, named after John Forbes Nash, Jr, is a fundamental
concept in game theory. The existence and computing of Nash
equilibria are two central problems in the theory of games. For
two-player zero-sum games, von Neumann proved the existence of
mixed-strategy equilibria using Brouwer Fixed Point Theorem. Nash
proved that if mixed strategies are allowed, then every game with
a finite number of players and strategies has at least one Nash
equilibrium (Nash, 1951). Although pure strategies are
conceptually simpler than mixed strategies, it is usually
difficult to guarantee the existence of a pure-strategy
equilibrium. However, it is shown that every finite potential game
possesses a pure-strategy Nash equilibrium (Monderer \& Shapley,
1996). The concept of potential game was proposed by
Rosenthal(1973). A game is said to be a potential game if it
admits a potential function. The incentive of all players to
change their strategy can be expressed by the difference in values
of the potential function. For a potential game, the set of
pure-strategy Nash equilibria can be found by searching the
maximal values of the potential function.
\par
An important problem is how to check whether a game is a potential
   game. Monderer
and Shapley (1996) first proposed necessary and sufficient
conditions for potential games. But it is required to verify all
the simple closed paths with length $4$ for any pair of players.
Then Hino (2011) gave an improved condition for detecting
potential games, which has a lower complexity than that of
Monderer and Shapley (1996) due to that only the adjacent pairs of
strategies of two players need to check. In Ui (2000), it is
proved that a game is potential if and only if the payoff
functions coincide with the Shapley value of a particular class of
cooperative games indexed by the set of strategy profiles. Game
decomposition is an important method for potential games
(Candogan, Menache, Ozdaglar, \& Parrilo, 2011; Hwang, \&
Rey-Bellet 2011; Sandholm, 2010) and some new necessary and
sufficient conditions for detecting potential games are obtained.
Sandholm (2010) established connections between his results and
that in Ui (2000). But the number of the obtained verification
equations is not the minimum. In Sandholm (2010), it is proved
that a finite game is a potential game if and only if, in each of
the component games, all active players have identical payoff
functions, and that in this case, the potential function can be
constructed.
\par
Recently, Cheng (2014) developed a novel method, based on the
semi-tensor product of matrices, to deal with games including
potential games, networked games and evolutionary games (Cheng,
2014; Cheng, Xu \& Qi, 2014; Cheng, Xu, He, Qi, 2014;  Cheng, He,
Qi, \& Xu, 2015; Guo, Wang, \& Li, 2013). In Cheng (2014), a
linear system, called potential equation, is proposed, and then it
is proved that the game is potential if and only if the potential
equation is solvable. With a solution of the potential equation,
the potential function can be directly calculated.
\par
A natural question is how to establish the connection between the
potential equation and the other criteria of potential games.
Moreover, an interesting problem is how to get the verification
equations with the minimum number. In this paper, we further
investigate the solvability of the potential equation. An
equivalence transformation is constructed to convert the augmented
matrix of the potential equation into the reduced row echelon
form. Based on this technique, some new necessary and sufficient
conditions for potential games are obtained. For potential games,
a new formula to calculate the potential functions is proposed.
Based on the obtained results, it is revealed the connection
between the potential equation and the results in Hino (2011) and
Sandholm (2010).
\par
\indent Throughout the paper, we denote the $k\times k$ identity
matrix by $I_k$, the $i$-th column of $I_k$ by $\delta^{i}_{k}$,
the $n$-dimensional column vector whose entries are all equal to 1
by $\mathbf{1}_k$, Kronecker product by $\otimes$ and the real
number field by $\mathbb{R}$.

\section{Preliminaries}
\begin{defn}
\label{defn1} (Monderer \& Shapley 1996) {\rm A {\it finite game}
is a triple $\mathcal{G}=(\mathcal{N}, \ \mathcal{S}, \
\mathcal{C})$, where
\\ \ (i) $\mathcal{N}=\{1,2,\cdots, n\}$ is the set of players; \\ \
(ii) $\mathcal{S}=\mathcal{S}_1\times \mathcal{S}_2\times \cdots
\times\mathcal{S}_n$ is the strategy set, where each
$\mathcal{S}_i=\{s^i_1, s^i_2,\cdots,s^i_{k_i}\}$ is the strategy
set of player $i$;
\\ \
(iii) $\mathcal{C}=\{c_1,c_2,\cdots, c_n\}$ is the set of payoff
functions, where every $c_i: \mathcal{S}\rightarrow \mathbb{R}$ is
the payoff function of player $i$.}
\end{defn}
Let $c^\mu_{i_1i_2\cdots i_n}=c_\mu(s^1_{i_1}, s^2_{i_2},\cdots,
s^n_{i_n})$ where $1\leq i_s\leq k_s$ and $s=1,2,\cdots,n$. Then
the finite game can be described by the arrays
\begin{equation}
\label{Cmu} C_\mu=\{c^\mu_{i_1i_2\cdots i_n}|\ 1\leq i_s\leq k_s,\
s=1,2,\cdots,n\}
\end{equation}
with $\mu=1,2,\cdots,n$. Particularly, for a 2-player game, the
$k_1\times k_2$ matrices $C_1=(c^1_{ij})$ and $C_2=(c^2_{ij})$ are
payoffs of players $1$ and $2$ respectively. Therefore, a 2-player
finite game is also called a {\it bi-matrix game}, which is
usually denoted by the simple notation $\mathcal{G}=(C_1,C_2)$.
\begin{defn}
\label{defn2}(Monderer \& Shapley 1996) {\rm A finite game
$\mathcal{G}=(\mathcal{N}, \ \mathcal{S}, \ \mathcal{C})$ is said
to be {\it potential} if there exists a function
$p:\mathcal{S}\rightarrow \mathbb{R}$, called the {\it potential
function}, such that $ \label{eq1}
c_i(x,s^{-i})-c_i(y,s^{-i})=p(x,s^{-i})-p(y,,s^{-i}) $ for all
$x,y\in \mathcal{S}_i$, $s^{-i}\in \mathcal{S}^{-i}$ and
$i=1,2,\cdots, n$, where
$\mathcal{S}^{-i}=\mathcal{S}_1\times\cdots\times
\mathcal{S}_{i-1}\times \mathcal{S}_{i+1}\times\cdots\times
\mathcal{S}_{n}$.}
\end{defn}
\begin{defn}
\label{defn3} \rm  \hspace{-0.2cm} ({\it Cheng, Qi, \& Li, 2011}).
Assume $A\in\mathbb{R}^{m\times n}$, $B\in\mathbb{R}^{p\times q}$.
Let $\alpha=\mathrm{lcm}(n,p)$ be the least common multiple of $n$
and $p$. The left semi-tensor product of $A$ and $B$ is defined as
$ A\ltimes B=(A\otimes I_{\frac{\alpha}{n}})(B\otimes
I_{\frac{\alpha}{p}})$.
\end{defn}
Since the left semi-tensor product is a generalization of the
traditional matrix product, the left semi-tensor product $A\ltimes
B$ can be directly written as $AB$. Identifying each strategy
$s^i_j$ with the logical vector $\delta_{k_i}^{j}$ for
$i=1,2,\cdots,n$ and $j=1,2,\cdots, k_i$, Cheng (2014) gave a new
expression of the payoff functions using the left semi-tensor
product.
\begin{lem} (Cheng, 2014)
Let $x_i\in \mathcal{S}_i$ be any strategy expressed in the form
of logical vectors. Then, for any payoff function $c_i$ of a
finite game $\mathcal{G}$ shown in Definition \ref{defn1}, there
exists a unique row vector $V^c_i\in \mathbb{R}^n$ such that
\begin{equation}
\label{eq7-0} c_i(x_1, x_2,\cdots, x_n)=V^c_i x_1x_2\cdots x_n,
\end{equation}
where $V^c_i$ is called the structure vector of $c_i$ and
$i=1,2,\cdots,n$.
\end{lem}
\begin{rem}
{\rm It is easy to see that $V^c_i$ is just the row vector
composed of the elements of $C_i$ in the lexicographic order (see
(\ref{Cmu})). Let
$C=[(V^c_1)^\mathrm{T},(V^c_2)^\mathrm{T},\cdots,(V^c_n)^\mathrm{T}]^\mathrm{T}$.
Then $C$ is just the {\it payoff matrix} of $\mathcal{G}$ proposed
by Cheng (2014).}
\end{rem}
Without loss of generality, we assume $k_i=k$ for all
$i=1,2,\cdots,n$. In Cheng (2014), the potential equation is
proposed as follows:
\begin{equation}
\label{eq7-1} \Psi \xi=b,
\end{equation}
where
\begin{equation}
\label{eq22}\Psi\!\!=\!\!\!\left[\!\!\begin{array}{ccccc}
             -\!\Psi_1 & \Psi_2  &  &  &  \\
             -\!\Psi_1 &  & \Psi_3 & & \\
             \vdots  &  &         & \ddots & \\
             -\!\Psi_1 &  &         &        & \Psi_n
             \end{array}\!\!\right]\!\!, \xi\!\!=
                   \!\!\left[\!\!\begin{array}{c}
                     \xi_1 \\ \xi_2 \\ \vdots \\ \xi_n \end{array}
                     \!\!\right]\!\!,
                     b=\!\!\left[\!\!\begin{array}{c}
                     (V_2^c\!-\!V_1^c)^\mathrm{T} \\ (V_3^c\!-\!V_1^c)^\mathrm{T} \\ \vdots \\ (V_n^c\!-\!V_1^c)^\mathrm{T} \end{array}
                     \!\!\right]
\end{equation}
and $\Psi_i=I_{k^{i-1}}\otimes \mathbf{1}_k\otimes I_{k^{n-i}}$
for each $i=1,2,\cdots,n$.
\begin{lem}(Cheng, 2014)
\label{lem5} A finite game $\mathcal{G}$ shown in Definition
\ref{defn1} is a potential game if and only if
 the potential equation (\ref{eq7-1})
has a solution $\xi$. Moreover, as (\ref{eq7-1}) holds, the
potential function $p$ can be calculated by
\begin{equation}
\label{eq8-1}
(V^p)^\mathrm{T}=(V^c_1)^\mathrm{T}-(\mathbf{1}_k\otimes
I_{k^{n-1}})\xi_1.
\end{equation}
\end{lem}
\section{Bi-matrix games}%
In this section, we consider the 2-player finite game
$\mathcal{G}=(C_1,\ C_2)$, where $C_i\in \mathbb{R}^{k_1\times
k_2}$ for $i=1,2$. In this special case, the coefficients of the
potential equation (\ref{eq7-1}) become
\begin{equation}
 \Psi= [-\mathbf{1}_{\!k_1}\!\otimes \!I_{\!k_2}\ \ \
                        I_{\!k_1}\!\otimes
                        \!\mathbf{1}_{\!k_2}],\ \
                        b=(V_2^c\!-\!V_1^c)^\mathrm{T}.
\end{equation}
Before the main results of this section, we first introduce a
basic property on Kronecker product.
\begin{lem}
\label{lem-Horn} (Horn, 1994) Let $\mathrm{V}_\mathrm{r}(X)$
denotes the vectorization of the matrix $X$ formed by stacking the
rows of X into a single column vector. Then
\begin{equation}
\label{eq7-2} \mathrm{V}_\mathrm{r}(ABC)=(A\otimes
C^\mathrm{T})\mathrm{V}_\mathrm{r}(B).
\end{equation}
In particular, we have
\begin{equation}
\label{eq7-3} \mathrm{V}_\mathrm{r}(AB)=(A\otimes
I)\mathrm{V}_\mathrm{r}(B).
\end{equation}
\end{lem}
\begin{thm}
\label{thm1} Set $B_k=[I_{k-1},\ -\mathbf{1}_{k-1}]\in
\mathbb{R}^{(k-1)\times k}$. The bi-matrix game $\mathcal{G}=(C_1,
\ C_2)$ is potential if and only if
\begin{equation}
\label{eq9} B_{k_1}(C_2-C_1)B_{k_2}^\mathrm{T}=0,
\end{equation}
where $C_1,\ C_2\in \mathbb{R}^{k_1\times k_2}$. Moreover, as
(\ref{eq9}) holds, the matrix form of each potential function is
\begin{equation}
\label{eq9-1} P=C_1+[0_{k_1\times (k_1-1)}\ \
\mathbf{1}_{k_1}](C_2-C_1)+\lambda\mathbf{1}_{k_1}\!\mathbf{1}_{k_2}^\mathrm{T},
\end{equation}
 where $\lambda\in \mathbb{R}$ is an arbitrary number.
\end{thm}
{\bf Proof.} Let $D_k=[I_{k-1},\ 0]\in \mathbb{R}^{(k-1)\times
k}$. Then it is easy to check that
\begin{equation}
\label{eq10}  B_kD_k^\mathrm{T}=I_{k-1},\ \
D_k\delta^k_k=B_k\mathbf{1}_{k}=0.
\end{equation}
Let
\begin{eqnarray}
\label{eq11}  E&\!=&\![-\delta_{k_1}^{k_1}\!\!\otimes \!I_{k_2},\
                  B_{k_1}^\mathrm{T}\!\!\otimes \!\delta_{k_2}^{k_2},\
                  B_{k_1}^\mathrm{T}\!\!\otimes \!B_{k_2}^\mathrm{T}]^\mathrm{T}\!\!\in \!\mathbb{R}^{k_1\!k_2\times k_1\!k_2}, \\
           F&\!=&\![-\mathbf{1}_{k_1}\!\otimes \!I_{k_2},\
                                  D_{k_1}^\mathrm{T}\!\otimes \!\mathbf{1}_{k_2},\
                                                  D_{k_1}^\mathrm{T}\!\otimes \!D_{k_2}^\mathrm{T}]\!\in \!\mathbb{R}^{k_1\!k_2\times k_1\!k_2}.
\end{eqnarray}
By (\ref{eq10}), a straightforward computation shows that
\begin{equation}
\label{eq12}  EF=I_{\!k_1\!k_2}.
\end{equation}
Eq. (\ref{eq12}) shows that $E$ is nonsingular. So the potential
equation is equivalent to
\begin{equation}
\label{eq13}  E\Psi \xi=Eb.
\end{equation}
With simple calculations, we have
\begin{eqnarray}
 E[\Psi \ \ b]&\!=&\!\left[\begin{array}{c}
 -(\delta_{k_1}^{k_1})^\mathrm{T}\!\!\otimes \!I_{\!k_2}\\
                  B_{\!k_1}\!\!\otimes \!(\delta_{k_2}^{k_2})^\mathrm{T}\\
                  B_{\!k_1}\!\!\otimes \!B_{\!k_2}\end{array}\right][-\mathbf{1}_{\!k_1}\!\otimes \!I_{\!k_2}\ \ \
                        I_{\!k_1}\!\!\otimes \!\mathbf{1}_{\!k_2}\ \ b]\nonumber\\
 &\!=&\!\left[\begin{array}{ccc}
 I_{\!k_2} & -(\delta_{k_1}^{k_1})^\mathrm{T}\!\!\otimes \!\mathbf{1}_{\!k_2} & -((\delta_{k_1}^{k_1})^\mathrm{T}\!\!\otimes \!I_{\!k_2})b\\
                0 &   B_{\!k_1} & (B_{\!k_1}\!\!\otimes \!(\delta_{k_2}^{k_2})^\mathrm{T})b\\
                 0 & 0 & (B_{\!k_1}\!\!\otimes \!B_{\!k_2})b\end{array}\right]\nonumber
\end{eqnarray}
\begin{eqnarray}
\label{eq14}
 &\!=&\!\left[\begin{array}{cccc}
   I_{\!k_2} & 0 &        -\mathbf{1}_{\!k_2} & -((\delta_{k_1}^{k_1})^\mathrm{T}\!\!\otimes \!I_{\!k_2})b\\
       0 &  I_{\!k_1-1} &    -\mathbf{1}_{\!k_1-1}  & (B_{\!k_1}\!\!\otimes \!(\delta_{k_2}^{k_2})^\mathrm{T})b\\
                 0 & 0 & 0 & (B_{\!k_1}\!\!\otimes \!B_{\!k_2})b \end{array}\right].
\end{eqnarray}
From (\ref{eq14}), it follows that (\ref{eq13}) is solvable if and only if
\begin{equation}
\label{eq14-1} (B_{\!k_1}\!\!\otimes \!B_{\!k_2})b=0,
\end{equation}
whose matrix form is just (\ref{eq9}) by (\ref{eq7-2}). As
(\ref{eq9}) holds, from (\ref{eq14}), we get
\begin{equation}
\label{eq15}  \xi_1=-((\delta_{k_1}^{k_1})^\mathrm{T}\!\!\otimes
\!I_{\!k_2})(V^c_2-V^c_1)^\mathrm{T}- c\mathbf{1}_{k_2},
\end{equation}
where $c$ is an arbitrary constant. Substituting (\ref{eq15}) into
(\ref{eq8-1}) yields
\begin{equation}
\label{eq16} (V^p)^\mathrm{T}\!\!=\!(V^c_1)^\mathrm{T}\!\!+
\!(\mathbf{1}_{\!k_1}(\delta_{k_1}^{k_1})^\mathrm{T}\!\otimes
\!I_{\!k_2})(V^c_2\!-\!V^c_1)^\mathrm{T} \!+\!c
\mathbf{1}_{\!k_1\!k_2}.
\end{equation}
Using (\ref{eq7-3}), from (\ref{eq16}), we get
\begin{equation}
\label{eq16-1}
P=C_1+\mathbf{1}_{\!k_1}(\delta_{k_1}^{k_1})^\mathrm{T}(C_2\!-\!C_1)
\!+\!c \mathbf{1}_{\!k_1}\mathbf{1}_{\!k_2}^\mathrm{T},
\end{equation}
which is just (\ref{eq9-1}). $\Box$
\begin{cor}
\label{cor8} Given a bi-matrix game $\mathcal{G}=(C_1, \ C_2)$, we
write the relative payoffs in the matrix form as
\begin{equation}
\label{eq17} R=(r_{ij})=C_2-C_1=\left[\begin{array}{cc}
                        R_1 & \eta \\
                         \zeta^\mathrm{T} & r_{\!k_1\!k_2}
                         \end{array}\right],
\end{equation}
where $R_1\in \mathbb{R}^{(k_1-1)\times (k_2-1)}$. Then
$\mathcal{G}$ is potential if and only if
\begin{equation}
\label{eq18} R_1-\mathbf{1}_{k_1-1}\zeta^\mathrm{T}-\eta
\mathbf{1}_{k_2-1}^\mathrm{T}+r_{\!k_1\!k_2}
\mathbf{1}_{k_1-1}\mathbf{1}_{k_2-1}^\mathrm{T}=0,
\end{equation}
namely,
\begin{equation}
\label{eq19} r_{ij}-r_{\!ik_2}-r_{\!k_1j}+r_{\!k_1\!k_2}=0
\end{equation}
for all $i=1,2,\cdots,k_1-1$ and $j=1,2,\cdots,k_2-1$.
\end{cor}
{\bf Proof}. Considering
\begin{eqnarray}
\label{eq20} B_{k_1}R B_{k_2}^\mathrm{T}&=&   [I_{k_1-1},\
-\mathbf{1}_{k_1-1}]
                                              \left[\begin{array}{cc}
                                               R_1 & \eta \\
                                                  \zeta^\mathrm{T} & r_{\!k_1\!k_2}
                                                \end{array}\right]
                                                            \left[\begin{array}{c}
                                                           I_{k_1-1}\\
                                                          -\mathbf{1}_{k_1-1}^\mathrm{T}
                                                            \end{array}\right]\nonumber \\
&=& \!R_1\!-\!\mathbf{1}_{\!k_1-1}\!\zeta^\mathrm{T}\!-\!\eta
\mathbf{1}_{\!k_2-1}^\mathrm{T}\!+r_{\!k_1\!k_2}
\mathbf{1}_{\!k_1-1}\!\mathbf{1}_{\!k_2-1}^\mathrm{T},
\end{eqnarray}
we get the corollary from Theorem \ref{thm1}. $\Box$
\par
The condition (\ref{eq19}) of Corollary \ref{cor8} is similar to
the condition proposed in Theorem 3 of Hino (2011). It should be
noted that Theorem 3 of Hino (2011) considers finite weighted
potential games. Here, for the convenience of comparing our result
with that in Hino (2011), we rewrite Theorem 3 of Hino (2011) for
the special case of 2-player potential games in the language of
relative payoff matrix as follows:
\begin{prop}(see Theorem 3 of Hino, 2011)
The bi-matrix game $\mathcal{G}=(C_1, \ C_2)$ is potential if and
only if
\begin{equation}
\label{eq20-0} r_{ij}-r_{\!i+1,j}-r_{\!i,j+1}+r_{\!i+1,j+1}=0
\end{equation}
for all $i=1,2,\cdots,k_1\!-\!1$ and $j=1,2,\cdots,k_2\!-\!1$.
\end{prop}
The original four-cycle condition proposed by Monderer \& Shapley
(1996) is rewritten as:
\begin{prop}(see Corollary 2.9 of Monderer \& Shapley, 1996)
The bi-matrix game $\mathcal{G}=(C_1, \ C_2)$ is potential if and
only if
\begin{equation}
\label{eq20-00} r_{ij}-r_{\!i'\!,j}-r_{\!i,j'}+r_{\!i'\!,j'}=0
\end{equation}
for all $i,i'=1,2,\cdots,k_1$ and $j,j'=1,2,\cdots,k_2$.
\end{prop}
Obviously, our condition (\ref{eq19}) is different from Hino's
condition (\ref{eq20-0}) and they have the same complexity, but
(\ref{eq20-00}) has a larger complexity (Hino, 2011).
\vspace{-1mm}
\begin{rem}
{\rm From (\ref{eq9}) or (\ref{eq14-1}), we see that, given the
strategy set for bi-matrix games, the set of all the relative
payoff matrices of potential bi-matrix games is a
$(k_1+k_2-1)$-dimensional subspace, which is isomorphic to
\begin{equation}
\label{eq20-1} \mathcal{P}=\{b\in \mathbb{R}^{k_1\!k_2}|\ \
(B_{k_1}\!\otimes \!B_{k_2})b=0\}.
\end{equation}
Here, we call $\mathcal{P}$ the {\it potential subspace}. If a
bi-matrix game $\mathcal{G}=(C_1,\ C_2)$ is not a potential game,
then one can use the orthogonal projection onto $\mathcal{P}$ to
yield corresponding potential games.}
\end{rem}
\vspace{-1mm} The basic result on orthogonal projection is stated
as follows: \vspace{-1mm}
\begin{lem}(see page 430 of Meyer, 2000)
\label{lem6} Consider a linear subspace of $\mathbb{R}^n$ as
follows:
\begin{equation}
\mathcal{X}=\{v\in \mathbb{R}^n|\ Bv=0\}.
\end{equation}
If $B$ has a full row rank, then the orthogonal projection of $u$
onto $\mathcal{X}$ is
\begin{equation}
\label{eq8-2}
\mathrm{Proj}_\mathcal{X}u=(I_n-B^\mathrm{T}(BB^\mathrm{T})^{-1}B)u.
\end{equation}
\end{lem}
\vspace{-1mm} Now we consider the orthogonal projection onto the
potential subspace. \vspace{-1mm} \vspace{-1mm}
\begin{lem}
\label{lem10} Consider a bi-matrix game $\mathcal{G}=(C_1, \
C_2)$, where $C_1,C_2\in \mathbb{R}^{\!k_1\times k_2}$. Denote the
relative payoff matrix by $R\!=\!(r_{ij})\!=\!C_2\!-\!C_1$ and let
$H_k\!=\!I_{\!k}\!-\!\frac{1}{k}\mathbf{1}_k\mathbf{1}_k^\mathrm{T}$.
Then \vspace{-1mm}
\begin{equation} \label{eq20-2}
\mathrm{Proj}_{\mathcal{P}}\mathrm{V_r}(R)=(I_{k_1\!k_2}-H_{k_1}\otimes
H_{k_2})\mathrm{V_r}(R).
\end{equation}
\end{lem}
\vspace{-1mm}
{\bf Proof}. Let $\tilde B=B_{k_1}\otimes B_{k_2}$.
By Lemma \ref{lem6}, we have
\begin{eqnarray}
\label{eq20-3}
&&\mathrm{Proj}_{\mathcal{P}}\mathrm{V_r}(R)\nonumber \\
&\!=& \!(I_{k_1\!k_2}\!-\!{\tilde B}^\mathrm{T}({\tilde B}{\tilde
B}^\mathrm{T})^{-1}{\tilde
B})\mathrm{V_r}(R) \nonumber \\
&\!=&\!(I_{\!k_1\!k_2}\!\!\!-\!\!({B}_{\!k_1}^\mathrm{T}\!\!({B}_{\!k_1}\!\!{B}_{\!k_1}^\mathrm{T}\!)^{-\!1}\!{B}_{\!k_1}\!)\!\otimes
\!\!({B}_{\!k_2}^\mathrm{T}\!({B}_{\!k_2}\!\!{B}^\mathrm{T}_{\!k_2}\!)^{-1}\!{B}_{\!k_2}\!)\!)\mathrm{V_{\!r}}\!(R).
\end{eqnarray}
A straightforward computation shows that
\begin{eqnarray}
\label{eq20-4}
&&{B}_{k}^\mathrm{T}({B}_{k}{B}_{k}^\mathrm{T})^{-1}{B}_{k}\nonumber \\
&=&\left[\begin{array}{c}
             I_{k-1}\\ -\mathbf{1}^\mathrm{T}_{k-1}
             \end{array}\right](I_{k-1}+\mathbf{1}_{k-1}\mathbf{1}_{k-1}^\mathrm{T})^{-1}[ I_{k-1}\ \ -\mathbf{1}_{k-1}]\nonumber \\
&=&\left[\begin{array}{c}
             I_{k-1}\\ -\mathbf{1}^\mathrm{T}_{k-1}
             \end{array}\right](I_{k-1}-\frac{1}{k}\mathbf{1}_{k-1}\mathbf{1}_{k-1}^\mathrm{T})[ I_{k-1}\ \
             -\mathbf{1}_{k-1}]\nonumber\\
&=&\left[\begin{array}{cc}
             I_{k-1}-\frac{1}{k}\mathbf{1}_{k-1}\mathbf{1}_{k-1}^\mathrm{T} &
                                           -\frac{1}{k}\mathbf{1}_{k-1}\\
       -\frac{1}{k}\mathbf{1}_{k-1}^\mathrm{T} &  \frac{k-1}{k}
             \end{array}\right]\nonumber  \\
&=& I_k-\frac{1}{k}\mathbf{1}_{k}\mathbf{1}_{k}^\mathrm{T}=H_k.
\end{eqnarray}
From (\ref{eq20-3}) and (\ref{eq20-4}), it follows that
(\ref{eq20-2}) holds. $\Box$
\begin{thm}
\label{thm11} Consider a bi-matrix game $\mathcal{G}=(C_1, \
C_2)$, where $C_1,C_2\in \mathbb{R}^{\!k_1\times k_2}$. Let the
relative payoff matrix be $R\!=\!(r_{ij})\!=\!C_2\!-\!C_1$. Then
the following statements are equivalent:
\\
{\rm (i)} $\mathcal{G}$ is a potential game;\\
{\rm (ii)} $H_{k_1}RH_{k_2}=0$, where $H_k=I_k-\frac{1}{k}\mathbf{1}_k\mathbf{1}_k^\mathrm{T}$;\\
{\rm (iii)}
$r_{ij}=r_{i\mathrm{-ave}}+r^{j\mathrm{-ave}}-r_\mathrm{\!ave}$
for all $i=1,2,\cdots,k_1$ and $j=1,2,\cdots,k_2$, where
\begin{eqnarray}
\label{eq21} \
&r_{i\mathrm{-ave}}\!=\!\frac{1}{k_2}\!\sum_{\mu=1}^{k_2}r_{i\mu},\
\
r^{j\mathrm{-ave}}\!=\!\frac{1}{k_1}\!\sum_{\lambda=1}^{k_1}r_{\lambda
j}, &
 \\
\label{eq21-0}
&r_\mathrm{\!ave}=\frac{1}{k_1k_2}\sum_{\lambda=1}^{k_1}\sum_{\mu=1}^{k_2}r_{\lambda
\mu}.&
\end{eqnarray}
\end{thm}
{\bf Proof.} Obviously, $\mathcal{G}$ is a potential game if and
only if
$\mathrm{Proj}_{\mathcal{P}}\mathrm{V_r}(R)=\mathrm{V_r}(R)$,
where $\mathcal{P}$ is the potential subspace. Further by Lemma
\ref{lem10}, we have that $\mathcal{G}$ is potential if and only
if $(H_{k_1}\otimes H_{k_2})\mathrm{V_r}(R)=0$, i.e. $H_kRH_k=0$.
Moreover, a straightforward calculation shows that
\begin{eqnarray}
\label{eq21-1} &&H_kRH_k \nonumber\\
&=&(I_{k_1}-\frac{1}{k_1}\mathbf{1}_{k_1}\mathbf{1}_{k_1}^\mathrm{T})
R(I_{k_2}-\frac{1}{{k_2}}\mathbf{1}_{k_2}\mathbf{1}_{k_2}^\mathrm{T}) \nonumber\\
&=&R\!-\!\frac{1}{k_1}\mathbf{1}_{k_1}\mathbf{1}_{k_1}^\mathrm{T}R
\!-\!\frac{1}{k_2}R\mathbf{1}_{k_2}\mathbf{1}_{k_2}^\mathrm{T}
\!+\!\frac{\mathbf{1}_{k_1}^\mathrm{T}R\mathbf{1}_{k_2}}{k_1k_2}\mathbf{1}_{k_1}\mathbf{1}_{k_2}^\mathrm{T}.
\end{eqnarray}
From (\ref{eq21})-(\ref{eq21-1}), the equivalence between (ii) and
(iii) follows. $\Box$
\begin{rem}
{\rm For the case of $k_1=k_2$, Sandholm (2010) obtained the
results of Theorem \ref{thm11} using the method of game
decomposition. A similar result can be seen in Proposition 2.14 of
Hwang \& Rey-Bellet (2011). But here, we get the results from the
potential equation. Therefore, we have established a connection
between the potential equation and the results obtained by
Sandholm (2010). From (\ref{eq21}) and (\ref{eq21-0}), we see that
$r_{i\mathrm{-ave}}$ is the average relative payoff for given
strategy $s^1_i\in \mathcal{S}_1$, $r^{j\mathrm{-ave}}$ is the
average relative payoff for given strategy $s^2_j\in
\mathcal{S}_2$ and $r^\mathrm{ave}$ is the average relative payoff
of all the strategies. Therefore, Theorem \ref{thm11} displays an
economic meaning of potential games.}
\end{rem}
\section{The General Potential Equation}
In this section, we consider the general potential equation for
multi-player games and give new detecting conditions for potential
games. \par Multiplying (\ref{eq7-1}) on the left by
\begin{equation}
\label{eq23} \left[\!\begin{array}{cc}
             0 &  1   \\
             -I_{n-1} & \mathbf{1}_{n-1}
\end{array}
             \!\right]\otimes I_{k^n},
\end{equation}
we get the equivalent equation
\begin{equation}
\label{eq24} \left[\!\!\begin{array}{ccccc}
             -\!\Psi_{\!1} &   &  &  & \Psi_n \\
                     & -\!\Psi_{\!2} &  & & \Psi_n\\
                     &         &  \ddots  & & \vdots \\
              &  &         &       -\!\Psi_{\!n-\!1} & \Psi_n
             \end{array}\!\!\right]
                   \!\!\left[\!\!\begin{array}{c}
                     \xi_1 \\ \xi_2 \\ \vdots \\ \xi_n \end{array} \!\!\right]\!\!=
                     \!\!\left[\!\!\begin{array}{c}
                     (V_n^c\!-\!V_1^c)^\mathrm{T} \\ (V_n^c\!-\!V_2^c)^\mathrm{T} \\ \vdots \\ (V_n^c\!-\!V_{n-1}^c)^\mathrm{T} \end{array}
                     \!\!\!\right].
\end{equation}
Construct nonsingular matrix $T=[T_1^\mathrm{T}\ \ T_2^\mathrm{T}\
\ T_3^\mathrm{T}]^\mathrm{T}$, where
\begin{equation}
\label{eq25} T_i=\left[\!\begin{array}{cccc}
             -T_{i1} &  &&   \\
                     & -T_{i2} &&\\
                     &&\ddots &\\
                     &&& -T_{i,n-1}
\end{array}
             \!\right] \ \
\end{equation}
with
\begin{eqnarray}
\label{eq26} T_{1j}&=&I_{k^{j-1}}\otimes (\delta_k^k)^\mathrm{T}\otimes I_{k^{n-j}}, \\
\label{eq27} T_{2j}&=&I_{k^{j-1}}\otimes B_k\otimes I_{k^{n-j-1}}\otimes (\delta_k^k)^\mathrm{T}, \\
\label{eq28} T_{3j}&=&I_{k^{j-1}}\otimes B_k\otimes I_{k^{n-j-1}}\otimes B_k
\end{eqnarray}
for all $j=1,2,\cdots,n$. It is easy to check that
\begin{eqnarray}
\label{eq29} T_{\!1\!j}\Psi_{\!j}&\!=&\!(I_{k^{j\!-1}}\!\!\otimes \!(\delta_k^k)^\mathrm{T}\!\!\otimes \!I_{\!k^{n\!-j}}\!)
(\!I_{\!k^{j\!-1}}\!\!\otimes \!\mathbf{1}_{\!k}\!\!\otimes \!I_{\!k^{n\!-j}}\!)\!=\!I_{k^{n\!-1}},\\
T_{\!1\!j}\Psi_{\!n}&\!=&\!(I_{k^{j\!-1}}\!\!\otimes \!(\delta_k^k)^\mathrm{T}\!\!\otimes \!I_{\!k^{n\!-j}}\!)
(\!I_{\!k^{n\!-1}}\!\!\otimes \!\mathbf{1}_{\!k})
\nonumber \\
&\!=&\!I_{k^{j\!-1}}\!\!\otimes \!\!(\delta_k^k)^\mathrm{T}\!\!\otimes \!\!I_{k^{n\!-j\!-1}}\otimes \mathbf{1}_k,\\
T_{\!2\!j}\!\Psi_{\!j}&\!=&\!\!(I_{k^{j\!-\!1}}\!\!\otimes \!B_k\!\!\otimes \!I_{\!k^{n\!-j\!-1}}\!\!\otimes \!\!(\delta_k^k)^{\!\mathrm{T}}\!)
(\!I_{\!k^{j\!-1}}\!\!\otimes \!\mathbf{1}_{\!k}\!\!\otimes \!I_{\!k^{n\!-j}}\!)\!=\!\!0,\\
T_{\!2\!j}\Psi_{\!n}&\!=&\!(I_{k^{j\!-1}}\!\!\otimes \!B_k\!\!\otimes \!I_{\!k^{n\!-j\!-1}}\otimes (\delta_k^k)^\mathrm{T}\!)
(\!I_{\!k^{n\!-1}}\!\!\otimes \!\mathbf{1}_{\!k})\nonumber \\
&=&I_{k^{j\!-1}}\!\!\otimes \!B_k\!\!\otimes \!I_{\!k^{n\!-j\!-1}},
\\
T_{\!3\!j}\!\Psi_{\!j}&\!=&\!\!(I_{k^{j\!-\!1}}\!\!\otimes \!B_k\!\!\otimes \!I_{\!k^{n\!-j\!-1}}\!\!\otimes \!\!B_k)
(\!I_{\!k^{j\!-1}}\!\!\otimes \!\mathbf{1}_{\!k}\!\!\otimes \!I_{\!k^{n\!-j}}\!)\!=\!\!0,\\
T_{\!3\!j}\Psi_{\!n}&\!=&\!(I_{k^{j\!-1}}\!\!\otimes \!B_k\!\!\otimes \!I_{\!k^{n\!-j\!-1}}\otimes \!\!B_k\!)
(\!I_{\!k^{n\!-1}}\!\!\otimes \!\mathbf{1}_{\!k})=0.
\end{eqnarray}
So, multiplying (\ref{eq24}) on the left by $T$ yields
\begin{equation}
\label{eq30} \left[\!\!\begin{array}{cccc}
             I_{\!k^{n\!-\!1}} &   &  & -(\delta_k^k)^\mathrm{T}\!\!\otimes \!\!I_{k^{n\!-2}}\!\!\otimes \!\!\mathbf{1}_k \\
                     & \!\!\!\ddots &   & \vdots\\
                     &         &  \!\!I_{\!k^{n\!-\!1}}  &  -I_{\!k^{n\!-2}}\!\otimes \!\!(\delta_k^k)^\mathrm{T}\!\!\otimes \!\!\mathbf{1}_k   \\
                     0& \!\!\cdots & \!\!0     &   \!-\!B_k \otimes I_{k^{n-2}}\\
                     0& \!\!\cdots & \!\!0     &   \!-\!I_k\otimes B_k \otimes I_{k^{n-3}}\\
                     \vdots & \!\!\ddots & \!\!\vdots   &     \vdots\\
                       0& \!\!\cdots & \!\!0    & \!-\!I_{k^{n-2}} \otimes B_k\\
                     0& \!\!\cdots & \!\!0    &  0
                               \end{array}\!\!\right]
                   \!\!\!\left[\!\!\begin{array}{c}
                     \xi_1 \\ \xi_2 \\ \vdots \\ \xi_n \end{array} \!\!\right]\!\!\!=
                     \!\!T\!\!\left[\!\!\begin{array}{c}
                     (V_n^c\!-\!V_1^c)^\mathrm{T} \\ (V_n^c\!-\!V_2^c)^\mathrm{T} \\ \vdots \\ (\!V_{\!n}^c\!-\!V_{\!n-\!1}^c\!)^{\!\mathrm{T}} \end{array}
                     \!\!\!\right],
\end{equation}
that is,
\begin{equation}
\label{eq31} \left[\!\begin{array}{cc}
             I_{\!(n-1)k^{n-1}} &  \Gamma \\
                    0 &\Phi\\
                   0  &  0
                     \end{array}\!\right]
                   \!\left[\!\begin{array}{c}
                    \tilde \xi \\ \xi_n
                      \end{array} \!\right]\!\!\!=
                     \left[\!\!\begin{array}{c}
                     T_1\tilde b \\ T_2\tilde b \\ T_3\tilde b
                      \end{array}
                   \! \right],
 \end{equation}
where $\tilde \xi=[\xi_1^\mathrm{T}, \cdots,
\xi_{n-1}^\mathrm{T}]^\mathrm{T}$, $\tilde
b=[V_n^c\!-\!V_1^c,\cdots, V_n^c\!-\!V_{n-1}^c]^\mathrm{T}$,
$\Phi\!=\![\Phi_1^\mathrm{T},\Phi_2^\mathrm{T},\cdots,\Phi_{n-1}^\mathrm{T}]^\mathrm{T}$,
$\Gamma=[\Gamma_1^\mathrm{T},\Gamma_2^\mathrm{T},\cdots,\Gamma_{n-1}^\mathrm{T}]^\mathrm{T}$
with
\begin{equation}
\label{eq31-1} \Phi_{\!i}\!=-\!I_{\!k^{i\!-\!1}}\!\otimes
\!B_k\!\otimes \!I_{\!k^{n\!-\!i\!-\!1}},\ \
\Gamma_{\!i}\!=\!-\!I_{\!k^{i\!-\!1}}\!\otimes
\!(\delta_{\!k}^k)^{\!\mathrm{T}}\!\otimes
\!I_{\!k^{n\!-\!1\!-\!i}}\!\otimes \!\mathbf{1}_{\!k}
 \end{equation}
for each $i=1,2,\cdots,n\!-\!1$. \par By
(\ref{eq31}), we get the proposition as follows:
\begin{prop}
\label{prop1}
The finite game $G$ is potential if and only if $T_3\tilde b=0$
and the linear equation
\begin{equation}
\label{eq32} \Phi \xi_n=T_2\tilde b
 \end{equation}
has a solution $\xi_n$.
\end{prop}
In the following, we consider (\ref{eq32}). Let
\begin{equation}
\label{eq33} S\!=\!\!\!\left[\!\!\begin{array}{c}
            S_1\\ S_2
            \end{array}
             \!\!\right]\!\!=\!\!\!\left[\!\!\begin{array}{cccccc}
             N_{11} & N_{12} & N_{13} &\cdots & N_{1,n\!-\!2} & N_{1,n\!-\!1}\\
             \hline
             M_{21} & L_{22} & L_{23} &\cdots & L_{2,n\!-\!2} & L_{2,n\!-\!1}\\
                    & M_{32} & L_{33} &\cdots & L_{3,n\!-\!2} & L_{3,n\!-\!1}\\
                    &        & \!\ddots & \ddots & \vdots  & \vdots \\
                    &        &        & \!\!\ddots & \ L_{n-2,n-2}& L_{n\!-\!2,n\!-\!1}       \\
                    &        &        &        & M_{n\!-\!1,n\!-\!2} &L_{n\!-\!1,n\!-\!1}
            \end{array}
             \!\!\right],
 \end{equation}
where
\begin{eqnarray}
\label{eq33-1} &&N_{ij}=D_{k^{n-i}}(I_{k^{j-i}}\otimes
D_k^\mathrm{T}\otimes
\mathbf{1}_{k^{n-j-1}}(\delta_{\!k^{n-j-1}}^{k^{n-j-1}})^\mathrm{T}),\\
&& L_{ij}=-I_{\!k^{i-2}}\!\otimes \!B_k \!\otimes \!N_{ij}, \\
&&M_{i,i-1}=I_{k^{i-2}(k-1)}\!\otimes \!B_{\!k^{n-i}}
 \end{eqnarray}
for all $i=1,2,\cdots,n-1$ and $j=i,i+1,\cdots,n-1$.
\par It is easy to check that $S$ is a square matrix with order
$(n-1)(k-1)k^{n-2}$. In order to prove $S$ is nonsingular, we
construct a square matrix as follows:
\begin{equation}
\label{eq33-1.1} U\!\!=\!\!\!\left[\!\!\begin{array}{ccccc}
             -\Phi_1D_{\!\!k^{n\!-\!1}}^\mathrm{T} & G_1 & \\
             -\Phi_2D_{\!\!k^{n\!-\!1}}^\mathrm{T} &  &  G_2  \\
                \vdots    &  & & \ddots & \\
                -\Phi_{\!n\!-\!2}D_{\!\!k^{n-1}}^\mathrm{T}    &  && & G_{n-2}\\
                -\Phi_{\!n\!-\!1}D_{\!\!k^{n\!-\!1}}^\mathrm{T}   &       &        &
                & 0
            \end{array}
             \!\!\right],
 \end{equation}
where $G_i=I_{\!k^{i-1}(\!k\!-\!1)}\!\!\otimes
\!\!D_{\!\!k^{n\!-\!i\!-\!1}}^\mathrm{T}$ for $i=1,2,\cdots, n-2$.

\begin{lem}
\label{lem-18} For matrices $S$ and $U$, we have
\begin{eqnarray}
\label{eq33-2} &&-\sum_{j=1}^{n-1}N_{1j}\Phi_j
\!=B_{\!k^{n\!-\!1}}, \\
\label{eq33-3}&&
M_{i,i-1}\Phi_{i-1}+\sum_{j=i}^{n-1}L_{ij}\Phi_j=0\ \
(i\!=\!2,3,\cdots\!, n\!-\!1\!),
\\
&& \label{eq33-4} SU=I_{(n-1)(k-1)k^{n-2}}.
 \end{eqnarray}
\end{lem}
{\bf Proof.} See Appendix. \par From (\ref{eq33-4}), it follows
that $S$ is nonsingular. Multiplying (\ref{eq32}) on the left by
$S$, we get an equivalent equation
\begin{equation}
\label{eq34} S\Phi \xi_n=ST_2\tilde b.
 \end{equation}
Let $S\Phi=[\Upsilon_1^\mathrm{T},
\Upsilon_2^\mathrm{T},\cdots,\Upsilon_{n-1}^\mathrm{T}]^\mathrm{T}$.
From (\ref{eq33-2}) and (\ref{eq33-3}), it follows that
$\Upsilon_1\!=\!B_{\!k^{n\!-\!1}}=[I_{\!k^{n\!-\!1}-1}\ \
-\mathbf{1}_{\!k^{n\!-\!1}-1}]$ and $\Upsilon_i=0$ for all
$i=2,3,\cdots, n-1$. Thus the linear equation (\ref{eq34}) is just
\begin{eqnarray}
\label{eq36-1} && [I_{\!k^{n\!-\!1}-1}\ \
 -\mathbf{1}_{\!k^{n\!-\!1}-1}]\xi_n=S_1T_2\tilde b,\\
\label{eq36-2}&& 0=S_2T_2\tilde b.
\end{eqnarray}
Then we get the following proposition.
\begin{prop}
\label{prop2}
The linear equation (\ref{eq32}) has a solution $\xi_n$ if and only if $S_2T_2\tilde b=0$.
\end{prop}
\begin{thm}
\label{thm20} Consider the finite game $\mathcal{G}=(\mathcal{N},
\mathcal{S}, \mathcal{C})$ described by Definition \ref{defn1}
with payoff functions in (\ref{eq7-0}). $\mathcal{G}$ is potential
if and only if
\begin{equation}
\label{eq38} \left[\begin{array}{c}
            S_2T_2\\ T_3
            \end{array}
            \right]\tilde b=0,
 \end{equation}
where $\tilde b=[V_n^c\!-\!V_1^c,\cdots,
V_n^c\!-\!V_{n-1}^c]^\mathrm{T}$ and the matrices $T_2$, $T_3$ and
$S_2$ are shown in (\ref{eq25}) and (\ref{eq33}). Moreover, as
(\ref{eq38}) holds, a potential function is described by
\begin{equation}
\label{eq39} p(x_1,\cdots, x_n)=V^px_1x_2\cdots x_n,
 \end{equation}
where
\begin{eqnarray}
\label{eq40} (V^p)^\mathrm{T}&=&(V_1^c)^\mathrm{T}+(\mathbf{1}_k(\delta_k^k)^\mathrm{T}\otimes I_{k^{n-1}})(V_n^c-V_1^c)^\mathrm{T}\nonumber \\
&&-\!\sum_{j=2}^{n-1}(\mathbf{1}_k(\delta_k^k\!)^\mathrm{T}\!\otimes
\!I_{\!k^{j-2}}\!\otimes \!D_{\!k}^\mathrm{T}B_{\!k}\!\otimes
\!\mathbf{1}_{k^{n-j}}\!(\delta_{k^{n-j}}^{\!k^{n-j}}\!)^\mathrm{T}\!)\nonumber \\
&&\ \ \ \ \cdot
(\!V_n^c\!\!-\!\!V_j^c\!)^\mathrm{T}\!+\!c\mathbf{1}_{\!k^n}.
 \end{eqnarray}
\end{thm}
{\bf Proof}. From Proposition \ref{prop1} and Proposition
\ref{prop2}, it follows that $\mathcal{G}$ is potential if and
only if (\ref{eq38}) holds. Now, we compute the potential
function. From (\ref{eq36-1}), we get the
\begin{equation}
\label{eq41} \xi_n=\left[\begin{array}{c}
                     S_1\\ 0
                     \end{array}\right]T_2\tilde b+c\mathbf{1}_{k^{n-1}},
 \end{equation}
where $c$ is an arbitrary constant. From (\ref{eq33}) and
(\ref{eq33-1}), it follows that
\begin{equation}
\label{eq41-1} S_1=D_{k^{n-1}}\tilde S_1,
\end{equation}
where $\tilde S_1=[\tilde S_{11}, \tilde S_{12}, \cdots,\tilde
S_{1,n-1}]$ with
\begin{equation} \label{eq41-2}\tilde
S_{1\!j}=I_{k^{j-1}}\otimes D_k^\mathrm{T}\otimes
\mathbf{1}_{k^{n-j-1}}(\delta_{\!k^{n-j-1}}^{k^{n-j-1}})^\mathrm{T}.
\end{equation}
Considering the last row of $\tilde S_{1}$ is $0$, by (\ref{eq41})
and (\ref{eq41-1}), we have
\begin{equation}
\label{eq41-3} \xi_n=\tilde S_1 T_2\tilde b+c\mathbf{1}_{k^{n-1}}.
 \end{equation}

By the first equation of (\ref{eq30}), we obtain that
\begin{eqnarray}
\label{eq42} \xi_1&\!=&\!-T_{\!11}(V_n^c\!-\!V_1^c)^\mathrm{T}+((\delta_k^k)^\mathrm{T}\!\!\otimes \!\!I_{k^{n\!-2}}\!\!\otimes \!\!\mathbf{1}_k)\xi_n\nonumber \\
&\!=&\!-T_{\!11}\!(V_n^c\!-\!\!V_1^c)^\mathrm{T}\!\!+\!(\!(\delta_k^k)^\mathrm{T}\!\!\otimes
\!\!I_{\!k^{n\!-2}}\!\!\otimes \!\!\mathbf{1}_{\!k})\!(\!\tilde
S_{\!1} T_{\!2}\tilde b\!+\!c\mathbf{1}_{\!k^{n\!-\!1}}\!).
 \end{eqnarray}
\vspace{-1mm}Substituting (\ref{eq42}) into (\ref{eq8-1}), we have
\vspace{-1mm}
\begin{eqnarray}
\label{eq43} (V^p)^\mathrm{T}&=&(V_1^c)^\mathrm{T}\!-\!(\mathbf{1}_k\!\otimes \!I_{k^{n-1}})\xi_1\nonumber \\
&=&(V_1^c)^\mathrm{T}\!+\!(\mathbf{1}_k\!\otimes
\!I_{k^{n-1}})T_{11}(V_n^c\!-\!V_1^c)^\mathrm{T}\nonumber \\
&& \!\!\!-\!(\!(\delta_k^k)^\mathrm{T}\!\!\otimes
\!\!I_{\!k^{n\!-2}}\!\!\otimes
\!\!\mathbf{1}_{\!k}\!)\!\sum_{j=1}^{n-1} \!\!\tilde
S_{1j}T_{2j}(V_n^c\!-\!\!V_j^c)^\mathrm{T}\!\!+\!c\mathbf{1}_{\!k^{n}}\!.
%
 \end{eqnarray}
\vspace{-1mm} Since $D_k\delta_k^k=0$, we have
\begin{eqnarray}
\label{eq44} &&((\delta_k^k)^\mathrm{T}\!\!\otimes
\!\!I_{\!k^{n\!-2}}\!\!\otimes \!\!\mathbf{1}_{\!k}\!)\tilde
S_{11}\nonumber\\
&=& ((\delta_k^k)^\mathrm{T}\!\!\otimes
\!\!I_{\!k^{n\!-2}}\!\!\otimes
\!\!\mathbf{1}_{\!k}\!)(D_k^\mathrm{T}\otimes
\mathbf{1}_{k^{n-2}}(\delta_{\!k^{n-2}}^{k^{n-2}})^\mathrm{T})=0.
 \end{eqnarray}
\vspace{-1mm}Substituting $T_{11}$, $T_{2j}$, $\tilde S_{1j}$ and
(\ref{eq44}) into (\ref{eq43}) yields (\ref{eq40}). \vspace{-1mm}
\begin{rem}
{\rm In Section 4 of Sandholm (2010), it is revealed that the
minimal number of
 linear equations to test potential games is
 $(n-1)k^n-nk^{n-1}+1$ for a $n$-player games with $k$ strategies.
However, for both the methods of Sandholm (2010) and Hino (2011),
the number of equalities to be verified is
$\frac{1}{2}n(n-1)k^{n-2}(k-1)^2$ (see the footnote of page 455 of
Sandholm (2010)), which is much greater than the minimal number.
Fortunately, using the potential equation in Cheng (2014), we get
the minimal number of equations described by (\ref{eq38}).}
\end{rem}
\vspace{-1mm}
\begin{thm}
\label{thm21} For the finite game $\mathcal{G}=(\mathcal{N},
\mathcal{S}, \mathcal{C})$ described by Definition \ref{defn1}
with payoff functions shown in (\ref{eq7-0}), the following
statements are equivalent:\\ {\rm (i)} $\mathcal{G}$ is
potential;\\ {\rm (ii)} equalities\vspace{-1mm}
\begin{eqnarray}
\label{eq45-1} && \!(I_{k^{i-1}}\otimes B_k\otimes
I_{k^{n-i-1}}\!\otimes \!B_k)(V_n^c-V_i^c)^\mathrm{T}=0,
\\
\label{eq45-2}  &&\!(I_{\!k^{i\!-\!1}}\!\!\otimes
\!B_{\!k}\!\otimes \!I_{\!k^{j\!-\!i\!-\!1}}\!\!\otimes
\!B_{\!k}\!\!\otimes \!I_{\!k^{n\!-\!j\!-\!1}}\!\!\otimes
\!(\delta_k^k)^\mathrm{T}\!)(V_j^c\!-\!\!V_i^c)^\mathrm{T}\!\!=\!0
\end{eqnarray}
hold for all $1\leq i<j\leq n-1$;
\\
{\rm (iii)} equalities\vspace{-1mm}
\begin{equation}
\label{eq46} (I_{\!k^{i-1}}\!\otimes \!B_{\!k}\!\otimes
\!I_{\!k^{j-i-1}}\!\otimes \!B_{\!k}\!\otimes
\!I_{\!k^{n-j}})(V_j^c-V_i^c)^\mathrm{T}\!=\!0
\end{equation}
 hold  for all
$1\leq i<j\leq n$.
\end{thm}
\vspace{-1mm}{\bf Proof}. (i)$\Rightarrow$(ii) Assume that
$\mathcal{G}$ is potential. Then, from Proposition \ref{prop1}, it
follows that $T_3\tilde b=0$ and the linear equation (\ref{eq32})
is solvable. It is easy to check that $T_3\tilde b=0$ is just
(\ref{eq45-1}). Moreover, (\ref{eq32}) implies that
\begin{equation} \label{eq47} (I_{\!k^{s\!-\!1}}\!\otimes\!
B_{\!k}\!\otimes
\!I_{\!k^{n\!-\!s\!-\!1}})\xi_n\!=\!(I_{\!k^{s\!-\!1}}\!\otimes
\!B_{\!k}\!\otimes \!I_{\!k^{n\!-\!s\!-\!1}}\!\otimes
\!(\!\delta_k^k)^\mathrm{T}\!)\!(V^c_n\!-\!V^c_s\!)^\mathrm{T}.
 \end{equation}
Letting $s=i$ in (\ref{eq47}) and multiplying (\ref{eq47}) on the
left by $I_{\!k^{j-1}}\otimes B_k\otimes\!I_{\!k^{n-j-1}}$ yield
\begin{eqnarray}
\label{eq48} &&(I_{\!k^{i-1}}\otimes B_k\otimes \!I_{\!k^{j-i+1}}\otimes B_k\otimes
\!I_{\!k^{n\!-\!j\!-\!1}})\xi_n\nonumber \\
&=&\!(I_{\!k^{i\!-\!1}}\!\!\otimes \!B_k\!\!\otimes \!I_{\!k^{j\!-\!i\!+\!1}}\!\!\otimes \!B_k\!\!\otimes
\!I_{\!k^{n\!-\!j\!-\!1}}\!\!\otimes \!(\delta_k^k)^\mathrm{T})(V^c_n\!-\!V^c_i)^\mathrm{T}\!.
 \end{eqnarray}
Similarly, letting $s=j$ in (\ref{eq47}) and multiplying
(\ref{eq47}) on the left by  $I_{\!k^{i-1}}\otimes
B_k\otimes\!I_{\!k^{n-i-1}}$ yield \vspace{-1mm}
\begin{eqnarray}
\label{eq49} &&(I_{\!k^{i\!-\!1}}\otimes B_{\!k}\otimes
\!I_{\!k^{j\!-i\!+1}}\otimes B_{\!k}\otimes
\!I_{\!k^{n\!-j\!-\!1}})\xi_n\nonumber \\
&=&\!(I_{\!k^{i\!-\!1}}\!\!\otimes \!B_k\!\!\otimes \!I_{\!k^{j\!-i\!+\!1}}\!\!\otimes \!B_k\!\!\otimes
\!I_{\!k^{n\!-j\!-1}}\!\!\otimes \!(\delta_k^k)^\mathrm{T})(V^c_n\!-\!V^c_j)^\mathrm{T}\!.
 \end{eqnarray}
Subtracting (\ref{eq49}) from (\ref{eq48}), we get (\ref{eq45-2}).
\\
(ii)$\Rightarrow$(iii) As $j=n$, (\ref{eq46}) is just (\ref{eq45-1}).
For the case of $1\leq i< j\leq n-1$, from (\ref{eq45-1}) and (\ref{eq45-2}), it follows that
\begin{eqnarray}
\label{eq50} &&\!(I_{\!k^{i-1}}\!\otimes \!B_{\!k}\!\otimes
\!I_{\!k^{j-i-1}}\!\otimes \!B_{\!k}\!\otimes
\!I_{\!k^{n-j}})(V_j^c-V_i^c)^\mathrm{T}\nonumber \\
&=&\!(\!I_{\!k^{i\!-\!1}}\!\!\otimes \!\!B_{\!k}\!\otimes
\!\!I_{\!k^{j\!-i\!-1}}\!\!\otimes \!\!B_{\!k}\!\otimes
\!\!I_{\!k^{n\!-j\!-1}}\!\!\otimes\!\!(D_{\!k}^\mathrm{T}\!B_{\!k}\!\!+\!\!\mathbf{1}_{\!k}(\delta_{\!k}^k)^{\!\mathrm{T}}\!)\!)(\!V_j^c\!\!-\!\!V_i^c\!)^\mathrm{T}\nonumber \\
 &=&\!(\!I_{\!k^{i\!-\!1}}\!\!\otimes \!\!B_{\!k}\!\otimes
\!\!I_{\!k^{j\!-i\!-1}}\!\!\otimes \!\!B_{\!k}\!\otimes
\!\!I_{\!k^{n\!-j\!-1}}\!\!\otimes\!\!D_{\!k}^\mathrm{T}\!B_{\!k}\!)(\!V_n^c\!\!-\!\!V_i^c\!)^\mathrm{T}\nonumber \\
&&-\!(\!I_{\!k^{i\!-\!1}}\!\!\otimes \!\!B_{\!k}\!\otimes
\!\!I_{\!k^{j\!-i\!-1}}\!\!\otimes \!\!B_{\!k}\!\otimes
\!\!I_{\!k^{n\!-j\!-1}}\!\!\otimes\!\!D_{\!k}^\mathrm{T}\!B_{\!k}\!)(\!V_n^c\!\!-\!\!V_j^c\!)^\mathrm{T}\nonumber \\
&&+\!(\!I_{\!k^{i\!-\!1}}\!\!\otimes \!\!B_{\!k}\!\otimes
\!\!I_{\!k^{j\!-i\!-1}}\!\!\otimes \!\!B_{\!k}\!\otimes
\!\!I_{\!k^{n\!-j\!-1}}\!+\!\mathbf{1}_{\!k}(\delta_{\!k}^k)^{\!\mathrm{T}}\!)(\!V_j^c\!\!-\!\!V_i^c\!)^\mathrm{T}\nonumber \\
&=&\!(\!I_{\!k^{j\!-\!1}}\!\!\otimes \!\!B_{\!k}\!\otimes
\!\!I_{\!k^{n\!-j\!-1}}\!\!\otimes D_k^\mathrm{T})
(\!I_{\!k^{i\!-\!1}}\!\!\otimes \!\!B_{\!k}\!\otimes
\!\!I_{\!k^{n\!-i\!-1}}\!\!\otimes \!B_k)(\!V_n^c\!\!-\!\!V_i^c\!)^\mathrm{T}\nonumber \\
&&\!\!-(\!I_{\!k^{i\!-\!1}}\!\!\otimes \!\!B_{\!k}\!\otimes
\!\!I_{\!k^{n\!-i\!-1}}\!\!\otimes
\!\!D_k^\mathrm{T}\!)(\!I_{\!k^{j\!-\!1}}\!\!\otimes
\!\!B_{\!k}\!\otimes
\!\!I_{\!k^{n\!-j\!-1}}\!\!\otimes \!\!B_k\!)(\!V_n^c\!\!-\!\!V_j^c\!)^{\!\mathrm{T}}\nonumber \\
&&\!\!-\!(I_{\!k^{n\!-\!2}}\!\!\otimes \!\!\mathbf{1}_{\!k})\!(I_{\!k^{i-1}}\!\!\otimes \!\!B_{\!k}\!\!\otimes
\!\!I_{\!k^{j\!-\!i\!-\!1}}\!\!\otimes\! \!B_{\!k}\!\otimes
\!I_{\!k^{n\!-\!j\!-\!1}}\!\!\otimes\!\!(\!\delta_{\!k}^k\!)^{\!\mathrm{T}}\!)(\!V_j^c\!-\!V_i^c\!)^{\!\mathrm{T}}\nonumber \\
&=&0.
 \end{eqnarray}
(iii)$\Rightarrow$(ii) Multiplying (\ref{eq46}) on the left by
$I_{k^{n-1}}\otimes (\delta_k^k)^\mathrm{T}$ yields
(\ref{eq45-2}). \par (ii)$\Rightarrow$(i) Since $T_3\tilde b=0$ is just (\ref{eq45-1}), by Theorem \ref{thm11}, we only need to check $S_2T_2=0$, i.e.
\begin{equation}
M_{i+1,i}T_{2i}\tilde b_i-\sum_{j=i+1}^{n-1}L_{i+1,j}T_{2j}\tilde b_j=0
 \end{equation}
for all $i\!=\!1,2,\!\cdots\!,n\!-\!2$. With simple calculations,
we have
\begin{eqnarray}
\label{eq51} &&L_{i+1,j}T_{2j}\nonumber \\
&=&(I_{\!k^{i\!-\!1}}\!\!\otimes \!\!B_k\!\!\otimes \!\!D_{k^{n\!-\!i\!-\!1}}(I_{\!k^{\!j\!-\!i\!-\!1}}\!\!\otimes \!\!D_k^\mathrm{T}\!\!\otimes \!\!\mathbf{1}_{k^{n\!-\!j\!-\!1}}(\delta_{\!k^{n\!-\!j\!-\!1}}^{k^{n\!-\!j\!-\!1}})^\mathrm{T}))\nonumber \\
&&\ \ \ \ \ \ \cdot (I_{\!k^{j\!-\!1}}\!\!\otimes \!\!B_k\otimes I_{k^{n-j-1}}\!\!\otimes \!\!(\delta_k^k)^\mathrm{T})\nonumber \\
&=&(\!I_{\!k^{i-1}(k-1)}\!\!\otimes
\!\!D_{\!k^{n\!-\!i\!-\!1}}\!)(\!I_{\!k^{j-2}(k-1)}\!\!\otimes
\!\!D_k^\mathrm{T}\!\!\otimes
\!\!\mathbf{1}_{k^{\!n\!-\!j\!-\!1}}(\!\delta_{\!k^{\!n\!-\!j\!-\!1}}^{k^{\!n\!-\!j\!-\!1}}\!)^\mathrm{T}\!)
\nonumber \\
&&\ \ \ \ \cdot (\!I_{\!k^{i\!-\!1}}\!\!\otimes \!B_{\!k}\!\otimes
\!I_{\!k^{j\!-\!i\!-\!1}}\!\!\otimes \!B_{\!k}\!\!\otimes
\!I_{\!k^{n\!-\!j\!-\!1}}\!\!\otimes
\!(\delta_k^k)^\mathrm{T}\!).
 \end{eqnarray}
From (\ref{eq45-2}) and (\ref{eq51}), it follows that
\begin{equation}
\label{eq51-1}
L_{i+1,j}T_{2j}\tilde b_j= L_{i+1,j}T_{2j}\tilde b_i.
 \end{equation}
Therefore, by (\ref{eq51-1}) and (\ref{eq33-3}), we have
\begin{eqnarray}
\label{eq51-3} &&M_{i+1,i}T_{2i}\tilde b_i\!-\!\sum_{j=i+1}^{n-1}L_{i+1,j}T_{2j}\tilde b_j\nonumber \\
&=&(M_{i+1,i}T_{2i}\!-\!\sum_{j=i+1}^{n-1}L_{i+1,j}T_{2j})\tilde b_i\nonumber \\
&=&((M_{i+1,i}\Phi_{i}\!-\!\sum_{j=i+1}^{n-1}L_{i+1,j}\Phi_{j})\otimes
(\delta_k^k)^\mathrm{T})\tilde b_i=0.
 \end{eqnarray}
$\Box$ \par Using the concept of multi-indexed matrix proposed by
Cheng (2012), we can simplify (\ref{eq46}). For details of
multi-indexed matrix, please refer Definition 1.1 and Definition
1.3 of Cheng (2012). Here, we only give an intuitive example.
Given a 4-dimensional data $$X\!\!=\!\{\!x_{i_1i_2i_3i_4}|\
i_1\!=\!1,\!2;\ i_2\!=\!1,\!2, \!3;\ i_3\!=\!1,\!2;\
i_4\!=\!1,\!2,\!3,\!4\!\},$$ we arrange $X$ into a matrix
$$
X^{14}_{23}=\left[\begin{array}{cccccc}
 x^{11}_{11} & x^{11}_{12} & x^{11}_{21} & x^{11}_{22} & x^{11}_{31} &
 x^{11}_{32}\\
x^{12}_{11} & x^{12}_{12} & x^{12}_{21} & x^{12}_{22} &
x^{12}_{31} &
 x^{12}_{32}\\
 x^{13}_{11} & x^{13}_{12} & x^{13}_{21} & x^{13}_{22} &
x^{13}_{31} &
 x^{13}_{32}\\
 x^{14}_{11} & x^{14}_{12} & x^{14}_{21} & x^{14}_{22} &
x^{14}_{31} &
 x^{14}_{32}\\
x^{21}_{11} & x^{21}_{12} & x^{21}_{21} & x^{21}_{22} &
x^{21}_{31} &
 x^{21}_{32}\\
x^{22}_{11} & x^{22}_{12} & x^{22}_{21} & x^{22}_{22} &
x^{22}_{31} &
 x^{22}_{32}\\
 x^{23}_{11} & x^{23}_{12} & x^{23}_{21} & x^{23}_{22} &
x^{23}_{31} &
 x^{23}_{32}\\
 x^{24}_{11} & x^{24}_{12} & x^{24}_{21} & x^{24}_{22} &
x^{24}_{31} &
 x^{24}_{32}\\
 \end{array}\right], $$
where $x^{i_1i_4}_{i_2i_3}=x_{i_1i_2i_3i_4}$, the multi-index of
the rows of $X^{14}_{23}$ is $i_1i_4$ and the multi-index of the
columns is $i_2i_3$. We usually say that $X^{14}_{23}$ is in the
order of $\mathrm{id}(i_1,i_4; 2,4)\times \mathrm{id}(i_2,i_3;
3,2)$.
\begin{lem}
\label{lem22} Suppose $Y=(A_1\otimes A_2\otimes\cdots\otimes
A_l)X$, where $X$ and $Y$ be column vectors and $A_i\in
\mathbb{R}^{m_i\times n_i}$ for each $i=1,2,\cdots, m$. Assume
that the elements of $X$ and $Y$ are in the order of $\
\mathrm{id}(j_1,\!\cdots\!,j_l;\ n_1,\!\cdots\!, n_l)$ and $\
\mathrm{id}(i_1,\!\cdots\!,i_l;\ m_1,\!\cdots\!, m_l)$
respectively. Then
$$
Y^{s_1\cdots s_t}_{r_1\cdots
r_{l-t}}\!=\!(A_{s_1}\!\otimes\!\cdots \!\otimes
\!A_{s_l})X^{s_1\cdots s_t}_{r_1\cdots
r_{l-t}}(A_{r_1}\!\otimes\!\cdots \!\otimes
\!A_{r_{l-t}})^\mathrm{T},
$$
where $\{s_1,\cdots, s_t, r_1,\cdots, r_{l-t}\}=\{1,2,\cdots,l\}$.
\end{lem}
By Lemma \ref{lem22} and (\ref{eq46}) of Theorem \ref{thm21}, we
get the following corollary.
\begin{cor}
Consider finite game $\mathcal{G}=(\mathcal{N}, \mathcal{S},
\mathcal{C})$ described by Definition \ref{defn1} with payoff
functions shown in (\ref{eq7-0}). Let $R^{j\rightarrow i}$ be the
multi-dimensional data of the relative payoffs with respect to
$j>i$, i.e. $R^{j\rightarrow i}=V^c_j-V^c_i$. Then $\mathcal{G}$
is potential if and only if
\begin{equation}
\label{eq52} (B_k\otimes B_k)(R^{j\rightarrow i})^{ij}_{12\cdots
\hat i\cdots \hat j\cdots n}=0
 \end{equation}
for all $1\leq i<j\leq n$, where a caret is used to denote missing
terms.
\end{cor}
\begin{rem}
{\rm Similar to Theorem \ref{thm11}, (\ref{eq52}) is equivalent to
\begin{equation}
\label{eq53} (H_k\otimes H_k)(R^{j\rightarrow i})^{ij}_{12\cdots
\hat i\cdots \hat j\cdots n}=0,
 \end{equation}
which is consistent with the result of Sandholm (2010). But our
result is derived from the potential equation. Thus we have
established a connection between the result of Cheng (2014) and
that of Sandholm (2010).}
\end{rem}
For players $i$ and $j$, arbitrarily given the strategies of the
other players, the payoffs of $i$ and $j$ admit a bi-matrix game,
which is called a {\it bi-matrix sub-game} of $\mathcal{G}$.
Obviously, the relative payoffs of each bi-matrix sub-game are
just lie in one column of some $(R^{j\rightarrow
i})^{ij}_{12\cdots \hat i\cdots \hat j\cdots n}$. Therefore, we
have the following corollary.
\begin{cor}
A finite game $\mathcal{G}$ is potential if and only if every
bi-matrix sub-game of $\mathcal{G}$ is potential.
\end{cor}
\section{An example}
\begin{exmp}
\label{exmp} {\rm Consider a finite game $\mathcal{G}$ with $n=3$,
$k=2$ and payoff matrix $C=(c^\mu_{i_1i_2i_3})$. Let
$$
R=\left[\begin{array}{cccccccc}
 r^1_{111} &  r^1_{112} & r^1_{121} & r^1_{122} & r^1_{211} &  r^1_{212} & r^1_{221} &
 r^1_{222}\\
r^2_{111} &  r^2_{112} & r^2_{121} & r^2_{122} & r^2_{211} &
r^2_{212} & r^2_{221} &
 r^2_{222}
\end{array}
 \right],
$$
where each $r^\mu_{ijk}=c^3_{ijk}-c^\mu_{ijk}$ for $\mu=1,2$.
 A computation shows that the
coefficient matrix of (\ref{eq38}) is
$$\left[\begin{array}{rrrrrrrrrrrrrrrr}
0 & 1 & 0 &-\!1 & 0 &-\!1 & 0 & 1 & 0 &-\!1 & 0 & 1 & 0 & 1 & 0 & -\!1\\
1 &-\!1 &0& 0& -\!1& 1 &0& 0& 0& 0& 0& 0& 0& 0& 0 &0\\
0&0& 1& -\!1&0&0& -\!1& 1&0 & 0& 0& 0& 0& 0& 0& 0\\
0&0&0& 0& 0& 0& 0& 0& 1& -\!1& -\!1& 1& 0& 0& 0& 0\\
0&0& 0& 0& 0& 0& 0&0&0& 0& 0& 0& 1& -\!1& -\!1& 1
\end{array}
\right].$$
Thus, by Theorem \ref{thm20}, the game is a potential
game if and only if
$$
\left\{\begin{array}{l}
r^1_{\!112}\!-\!r^1_{\!122}\!-\!r^1_{\!212}\!+\!r^1_{\!222}\!-\!r^2_{\!112}\!+\!r^2_{\!122}\!+\!r^2_{\!212}\!-\!r^2_{\!222}\!=\!0,\\
r^1_{\!111}-r^1_{\!112}-r^1_{\!211}+r^1_{\!212}=0,\\
r^1_{\!121}-r^1_{\!122}-r^1_{\!221}+r^1_{\!222}=0,\\
r^2_{\!111}-r^2_{\!112}-r^2_{\!211}+r^2_{\!212}=0,\\
r^2_{\!121}-r^2_{\!122}-r^2_{\!221}+r^2_{\!222}=0,
\end{array}\right.
$$
which has the minimal number of equations for detecting whether
$\mathcal{G}$ is potential.}

\end{exmp}
\section{Conclusions}
For detecting whether a finite game is potential, new necessary
and sufficient conditions have been obtained by investigating the
potential equations. The number of the obtained verification
equalities is minimal. The connections between the potential
equations and the existing results on potential games have been
revealed. It has been shown that a finite game is potential if and
only if its every bi-matrix sub-game is potential. In the future
work, we will use the potential equation to investigate
near-potential games, networked game and so on.
%



\begin{thebibliography}{00}
\bibitem{Candogan} Candogan O., Menache, I., Ozdaglar, A., \& Parrilo, P. A. (2011).
Flows and Decompositions of Games: Harmonic and Potential Games.
{\it Mathematics of Operations Research}, 36(3), 474-503.

\bibitem{Cheng_book} Cheng, D., Qi, H., \& Zhao, Y. (2012).
{\it An introduction to semi-tensor product of matrices and its
applications}. Singapore: World Scientific.

\bibitem{Cheng1} Cheng, D. (2014). On finite potential games. {\it Automatica}, 50(7): 1793-1801.

\bibitem{Cheng2} Cheng, D., Xu, T., \& Qi, H. (2014). Evolutionarily stable strategy of networked
evolutionary games. {\it IEEE Transactions on Neural Networks and
Learning Systems}, 25(7), 1335-1345.

\bibitem{Cheng3} Cheng, D., Xu, T., He, F., \& Qi, H. (2014). On dynamics and Nash
equilibriums of networked games, {\it IEEE/CAA Journal of
Automatica Sinica}, 1(1), 10-18.

\bibitem{Cheng4} Cheng, D., He, F., Qi, H., \& Xu, T. (2015). Modeling, analysis and control of
networked evolutionary games. {\it IEEE Transactions on Automatic
Control}, DOI: 10.1109/TAC.2015.2404471.

\bibitem{Guo} Guo, P., Wang, Y., \& Li, H. (2013). Algebraic formulation and
strategy optimization for a class of evolutionary networked games
via semi-tensor method. {\it Automatica}, 49(11), 3384-3389.

\bibitem{Hino} Hino Y. (2011). An improved algorithm for detecting potential
games. {\it International Journal of Game Theory}, 40(1), 199-205.

\bibitem{Horn} Horn, R. A., \& Johnson, C. R. {\it Topics in matrix
analysis}. UK: Cambridge University Press, 1994.

\bibitem{Hwang} Hwang, S.-H. \& Rey-Bellet L. (2011). Decompositions of two player games:
potential, zero-sum, and stable games. arXiv:1106.3552v2.

\bibitem{Meyer} Meyer, C. C., (2000). {\it Matrix analysis and applied linear
algebra}. USA: Society for Industrial and Applied Mathematics.

\bibitem{Nash} Nash, J. (1951). Non-cooperative games. {\it Annals of Mathematics},
54, 286-295.

\bibitem{Sandholm} Sandholm, W. H. (2010). Decompositions and potentials for normal form
games. {\it Games and Economic Behavior}, 70, 446-456.

\bibitem{Shapley} Monderer, D. \& Shapley L. S. (1996). Potential Games. {\it Games
and Economic Behavior}, 14, 124-143.

\bibitem{Ui} Ui, T. (2000). A shapley value representation of potential games.
{\it Games and Economic Behavior}, 31, 121-135.

\bibitem{Neumann} von Neumann, J. \& Morgenstern, O.
(1953). {\it Theory of Games and Economic Behavior}. USA:
Princeton University Press.




\end{thebibliography}


%
\section*{Appendix: ~~Proof of Lemma \ref{lem-18}}
By the basic fact
$D_k^\mathrm{T}B_k=I_k-\mathbf{1}_k(\delta_k^k)^\mathrm{T}$, we
have \vspace{-0.5mm}
\begin{eqnarray}
\label{eqA1} &\!&\!-\sum_{j=1}^{n-1}N_{1j}\Phi_j\nonumber\\
&\!=&\!\sum_{j=1}^{n-1}\!D_{\!k^{n\!-\!1}}(I_{\!k^{j\!-\!1}}\!\otimes
\!D_k^\mathrm{T}B_k\otimes
\mathbf{1}_{\!k^{n\!-\!j\!-\!1}}(\delta_{\!k^{n\!-\!j\!-\!1}}^{k^{n\!-\!j-\!1}})^\mathrm{T})\nonumber\\
&\!=&\!D_{\!k^{n\!-\!1}}\!\sum_{j=1}^{n-1}I_{\!k^{j\!-\!1}}\!\otimes
\!(I_k-\mathbf{1}_k(\delta_k^k)^\mathrm{T})\otimes
\mathbf{1}_{\!k^{n\!-\!j\!-\!1}}(\delta_{\!k^{n\!-\!j\!-\!1}}^{k^{n\!-\!j-\!1}})^\mathrm{T}\nonumber\\
&\!=&\!D_{\!k^{n\!-\!1}}\!(\sum_{j=1}^{n-1}\!\!I_{k^j}\!\!\otimes
\!\mathbf{1}_{\!k^{n\!-\!j\!-\!1}}\!(\delta_{\!k^{n\!-\!j\!-\!1}}^{k^{n\!-\!j-\!1
}}\!)^\mathrm{T}\!\!-\!\!
\!\!\sum_{j=1}^{n-1}\!\!I_{k^{j\!-\!1}}\!\!\otimes
\!\!\mathbf{1}_{\!k^{n\!-\!j}}\!(\delta_{\!k^{n\!-\!j}}^{k^{n\!-\!j}}\!)^\mathrm{T}\!)\nonumber\\
&\!=&\!D_{\!k^{n\!-\!1}}(I_{k^{n-1}}-\mathbf{1}_{\!k^{n\!-\!1}}\!(\delta_{\!k^{n\!-\!1}}^{k^{n\!-\!1}})^\mathrm{T})\nonumber\\
&\!=&\!D_{\!k^{n\!-\!1}}D_{\!k^{n\!-\!1}}^\mathrm{T}B_{\!k^{n\!-\!1}}\nonumber
\\
&\!=&\!B_{\!k^{n\!-\!1}}=[I_{\!k^{n\!-\!1}-1}\ \
-\mathbf{1}_{\!k^{n\!-\!1}-1}].
 \end{eqnarray}
For every $i=2,3,\cdots, n-1$, similarly, we have
\vspace{-0.5mm}
\begin{eqnarray}
\label{eqA2} &\!&\!-M_{i,i-1}\Phi_{i-1}-\sum_{j=i}^{n-1}L_{ij}\Phi_j\nonumber\\
&\!=&\!I_{\!k^{i\!-\!2}}\!\otimes \!B_k\otimes B_{\!k^{n\!-\!i}}-
\sum_{j=i}^{n-1}I_{\!k^{i\!-\!2}}\otimes B_k\otimes
(\!D_{\!k^{n\!-\!i}}\nonumber\\
&&\hspace{2cm}  \cdot(I_{\!k^{j\!-\!i}}\!\otimes
\!D_k^\mathrm{T}B_k\otimes
\mathbf{1}_{\!k^{n\!-\!j\!-\!1}}(\delta_{\!k^{n\!-\!j\!-\!1}}^{k^{n\!-\!j-\!1}})^\mathrm{T}))\nonumber\\
&\!=&\!I_{\!k^{i\!-\!2}}\!\otimes \!B_k\otimes B_{\!k^{n\!-\!i}}-
\sum_{j=i}^{n-1}I_{\!k^{i\!-\!2}}\otimes B_k\otimes
(\!D_{\!k^{n\!-\!i}}\nonumber\\
&&\hspace{1cm}  \cdot(I_{\!k^{j\!-\!i}}\!\otimes
\!(I_k-\mathbf{1}_k(\delta_k^k)^\mathrm{T}) \otimes
\mathbf{1}_{\!k^{n\!-\!j\!-\!1}}(\delta_{\!k^{n\!-\!j\!-\!1}}^{k^{n\!-\!j-\!1}})^\mathrm{T}))\nonumber\\
&\!=&\!I_{\!k^{i\!-\!2}}\!\!\otimes \!\!B_k\!\!\otimes
\!\!B_{\!k^{n\!-\!i}}\!- \!I_{\!k^{i\!-\!2}}\!\!\otimes
\!\!B_k\!\!\otimes
\!\!D_{\!k^{n\!-\!i}}\!(I_{\!k^{n\!-\!i}}\!-\!\mathbf{1}_{\!k^{n\!-\!1}}\!(\delta_{\!k^{n\!-\!i}}^{k^{n-\!i}}\!)^{\!\mathrm{T}}\!)
\nonumber\\
&\!=&\!I_{\!k^{i\!-\!2}}\!\!\otimes \!\!B_k\!\!\otimes
\!\!B_{\!k^{n\!-\!i}}\!- \!I_{\!k^{i\!-\!2}}\!\!\otimes
\!\!B_k\!\!\otimes
\!\!D_{\!k^{n\!-\!i}}D_{\!k^{n\!-\!i}}^\mathrm{T}B_{\!k^{n\!-\!i}}\nonumber\\
&\!=&\!0.
 \end{eqnarray}
Moreover, by the fact that $B_kD_k^{\mathrm{T}}=I_{k-1}$ we have
\vspace{-0.5mm}\begin{eqnarray} \label{eqA3}
M_{i,i-1}G_{i-1}&=&(I_{k^{i-2}(k-1)}\!\otimes
\!B_{\!k^{n-i}})(I_{k^{i-2}(k-1)}\!\otimes
\!D_{\!k^{n-i}}^\mathrm{T})\nonumber \\
&=&I_{k^{i-2}(k-1)(k^{n-i}-1)}.
 \end{eqnarray}
\vspace{-0.5mm}From $D_k\delta_k^k=0$, it follows that
\vspace{-0.5mm}\begin{eqnarray} \label{eqA4}
L_{ij}G_{j}&=&-(I_{\!k^{i-2}}\!\otimes \!B_k \!\otimes
\!N_{ij})(I_{k^{j-1}(k-1)}\!\otimes
\!D_{\!k^{n-j-1}}^\mathrm{T})\nonumber \\
&=&-I_{\!k^{i-2}}\!\otimes \!B_k \!\otimes
\!N_{ij}(I_{k^{j-i}(k-1)}\!\otimes
\!D_{\!k^{n-j-1}}^\mathrm{T})\nonumber \\
&=&-I_{\!k^{i-2}}\!\otimes \!B_k \!\otimes
\!D_{\!k^{n-i}}(I_{k^{j-i}}\!\otimes D_{\!k}^\mathrm{T}\nonumber
\\
&&\otimes
\mathbf{1}_{k^{n-j-1}}(\delta_{k^{n-j-1}}^{k^{n-j-1}})^\mathrm{T})(I_{k^{j-i+1}}\!\otimes
\!D_{\!k^{n-j-1}}^\mathrm{T})\nonumber \\
&=&-I_{\!k^{i-2}}\!\otimes \!B_k \!\otimes
\!D_{\!k^{n-i}}(I_{k^{j-i}}\!\otimes D_{\!k}^\mathrm{T}\nonumber
\\
&&\otimes
\mathbf{1}_{k^{n-j-1}}(D_{\!k^{n-j-1}}\delta_{k^{n-j-1}}^{k^{n-j-1}})^\mathrm{T})\nonumber \\
&=&0.
 \end{eqnarray}
From (\ref{eqA1})-(\ref{eqA4}), we get (\ref{eq33-4}).

\end{document}